\documentclass{article}

\usepackage{arxiv}

\usepackage[utf8]{inputenc} 
\usepackage[T1]{fontenc}    
\usepackage{hyperref}       
\usepackage{url}            
\usepackage{booktabs,makecell,multirow}       
\usepackage{amsfonts}       
\usepackage{nicefrac}       
\usepackage{microtype}      
\usepackage{lipsum}
\usepackage{graphicx}
\graphicspath{ {./images/} }
\usepackage{natbib}
\usepackage{amsmath,amssymb,amsthm}
\title{Forecasting hospital discharges for respiratory conditions in Costa Rica using climate and pollution data.}

\author{
 Shu Wei Chou-Chen \\
  Centro de Investigación en Matematica Pura y Aplicada\\
  Escuela de Estadística\\
  Universidad de Costa Rica\\
  Costa Rica
  \texttt{shuwei.chou@ucr.ac.cr} \\
   \And
 Luis A. Barboza \\
  Centro de Investigación en Matematica Pura y Aplicada\\
  Escuela de Matemática\\
  Universidad de Costa Rica\\
  Costa Rica
  \texttt{luisalberto.barboza@ucr.ac.cr} \\
}

\begin{document}
\maketitle
\begin{abstract}
Respiratory diseases represent one of the most significant economic burdens on healthcare systems worldwide. The variation in the increasing number of cases depends greatly on climatic seasonal effects, socioeconomic factors, and pollution. Therefore, understanding these variations and obtaining precise forecasts allows health authorities to make correct decisions regarding the allocation of limited economic and human resources. This study aims to model and forecast weekly hospitalizations due to respiratory conditions in seven regional hospitals in Costa Rica using four statistical learning techniques (Random Forest, XGboost, Facebook’s Prophet forecasting model, and an ensemble method combining the above methods), along with 22 climate change indices and aerosol optical depth as an indicator of pollution. Models are trained using data from 2000 to 2018 and are evaluated using data from 2019 as testing data. Reliable predictions are obtained for each of the seven regional hospitals.
\end{abstract}


\section{Introduction}

Respiratory diseases are conditions that affect the organs and tissues within the lungs and airway systems, leading to difficulty in breathing. These illnesses are currently the leading contributors to the global burden of diseases when assessed through disability-adjusted life-years. Moreover, the rising healthcare costs associated with these diseases are placing a growing strain on the economies of nations worldwide, specifically the five major conditions, known as the ``big five'', which are chronic obstructive pulmonary disease (COPD), asthma, acute respiratory infections, tuberculosis, and lung cancer \cite{FIRS2014}. To maintain a balanced health economy, it is crucial to understand the distribution and the incidence of these diseases to optimize healthcare systems, ensure efficient resource allocation, and enhance healthcare access and quality.

Extensive research indicates that climate influences respiratory health, resulting in higher rates of hospitalization and mortality in varying seasons \citep[see e.g.,][]{chen2022,lin2009,pedder2021,tran2022,tsangari2016}. 
Furthermore, it is well known that human activities significantly impact the short-term and long-term climate by releasing greenhouse gases and other pollutants into the atmosphere, leading to more extreme weather events in recent years. As a consequence, researchers worldwide have been studying the effects of pollution and climatological factors on human health in different regions of the world \cite{Guo2016,Kan2012,Liu2019,Lou2019,Miraglia2005}.

Specifically with respiratory diseases, climate factors are found to be associated with extreme weather \cite{makrufardi2023,tran2022,tsangari2016}, and pollution, such as particular matter (PM) \cite{Hamra2014,pedder2021}. Airborne particulate matter (PM) is a complex mixture of different chemical species originating from numerous sources. These particles can be a product of combustion, suspension of soil materials, suspension of substances from the sea, and can also be formed by chemical reactions in the atmosphere\cite{colbeck2010}. Furthermore, elevated concentrations of PM not only substantially heighten health risks but also contribute significantly to the challenges associated with climate change. Therefore, the monitoring of PM is essential for comprehending and mitigating air pollution, aiming to enhance public health. Nevertheless, the measurement of PM presents challenges in developing countries due to the constrained monitoring infrastructure, resulting in gaps in data analysis both temporally and spatially.

On the other hand, an aerosol is defined as a stable suspension of solid and liquid particles in gas. It is possible to obtain related measurements with the facilitation of remote sensing techniques. Aerosol optical depth (AOD), a satellite-derived metric that quantifies the presence of aerosols across the entire atmospheric column, is a good option. Several studies around the world have found that the driving factors of AOD are related to urban and economic development, agricultural activities, industrialization, landscape aggregation, and regional transportation \citep[see e.g.][]{provencal2017,zhang2018}. Moreover, AOD is found to be associated with socio-economic factors, such as GDP, industry, and vehicle density \cite{li2014}.

Costa Rica, a country with an area of 51,179 $Km^2$, is located in Central America with an estimated population of approximately 5,003,402 inhabitants in 2018. It is a stable democratic country with a socioeconomic development model based on the opening and economic liberalization, upholding human rights and ensuring universal basic access to essential goods and services, such as education and health system while preserving extensive natural resources. \cite{OPS2019}. 

Although Costa Rica is renowned for its biodiversity, with most environmental policies focusing on the protection and conservation of natural resources, unregulated land use has been observed, resulting in significant environmental impacts. This issue is not limited to the Great Metropolitan Area, the country's most urban and densely populated region, but is also prevalent in other parts of the national territory. Consequently, this form of human development can intensify vulnerability to natural disasters and extreme weather conditions\cite{OPS2019}.

Regarding to its health system, since the 1940s Costa Rica prompted its national health system and social investment by creating the Costa Rican Social Security Fund (CCSS, from its Spanish acronym \textit{Caja Costarricense de Seguro Social}).
The basis of the health system in Costa Rica is universal and public, which covers 95\% of the Costa Rican population \cite{Baird2023,OPS2019}. The CCSS carries all health care functions in the country and divides the territory into 7 Health Regions (\textit{región sanitaria} in Spanish), and at the same time, it is divided into 104 Health Areas (AS, from its Spanish acronym \textit{áreas de salud}), and then, 1045 Basic Team for Comprehensive Health Care (EBAIS, from its Spanish acronym \textit{Equipos Básicos de Atención Integral de Salud}). Each AS caters to a population ranging from 15,000 to 40,000 individuals in rural areas and 30,000 to 60,000 individuals in urban areas. Subsequently, these ASs are subdivided into 1,045 segments, each overseen by an EBAIS. Each segment, in turn, serves approximately 4,000 people.

The health service in Costa Rica operates on three distinct levels of attention, interconnected through referral and counter-referral mechanisms. The initial level of service serves as the primary entry point to the healthcare system, providing essential services within each EBAIS at the community level. Additionally, there are two possibilities in which a person can access the health system: 1) an insured person can be attended by a private doctor who is registered by the CCSS and access reference to more specialized attention, and 2) an insured person is attended by a doctor hired by the company.
The second level aims to provide support to the first level through a network of 10 main clinics, 13 peripheral hospitals, and 7 regional hospitals, through the provision of specialized outpatient services, outpatient and inpatient interventions, hospitalization, and medical-surgical treatment of medical specialties, and some others of high population demand, such as dermatology, urology and ophthalmology, all of low complexity. 
Finally, the third level includes outpatient and inpatient services of greater complexity and specialization that require high technology and level of specialization. It is provided through 3 national hospitals and 6 specialized hospitals; the area of influence of this level covers the territory of several provinces.

Due to this health administrative flow, the accurate diagnostics of respiratory conditions that require hospitalization are registered in hospital discharge events (second level), instead of hospital entrance (first level). A a consequence, the limitation of this data lies in the presence of a lagged and unknown effect, necessitating consideration in the analysis. Beyond the delayed impact associated with hospital admissions and discharges, there is also a temporal lag effect pertaining to climate and pollution variables.

In existing literature, most studies are linked to statistical association between climate and hospitalizations \citep[see e.g.,][]{chen2022,lin2009,pedder2021,tran2022,tsangari2016}. 
Additionally, some of them concentrate on forecasting hospitalizations specifically within emergency departments and respiratory diseases \cite[see e.g.][]{Peng2020,Souza2016,soyiri2013}.
Our main goal is to predict hospital discharges due to respiratory diseases at the secondary healthcare level of the system, in the hope of enhancing the country's healthcare economy. Specifically, we analyze and predict hospital discharges in 7 regions to enable health authorities to make accurate predictions and informed budget decisions in the near future. To the best of our knowledge, this approach is the first predictive study undertaken in the region with these distinctive characteristics.

This paper is divided as follows: Section \ref{sec:methods} describes the data and the implemented methodologies. Section \ref{sec:results} presents the analysis and forecasting results. Finally, section \ref{sec:discussion} discusses the prediction performance, limitations, and future work.

\section{Materials and methods}
\label{sec:methods}
\subsection{Data}

This section provides an overview of the primary datasets used in our study, focusing on respiratory hospitalization and input data comprising climate, pollution, and lagged hospital discharge.

\subsubsection{Respiratory hospitalization}

\begin{figure}[htp!]
    \centering
    \includegraphics[scale = 0.75]{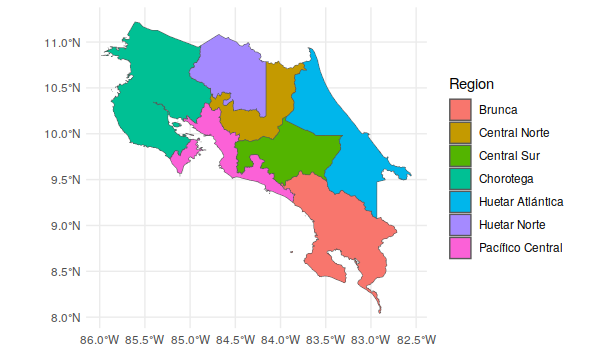}
    \caption{Health administrative division of Costa Rica.}
    \label{fig:mapa}
\end{figure}

Weekly data on hospital discharges due to all respiratory diseases (J00-J99 using ICD10 \cite{ICD10}) from 2000 to 2019 for seven regions (Brunca, Central Norte, Central Sur, Chorotega, Huetar Atlántica, Huetar Norte and Pacífico Central), were obtained from the Health Statistics Area of the CCSS. The choice of using 'discharge' instead of 'hospital entrance' is based on the accurate registration of diagnostics at the second level of the health system in Costa Rica. According to the official statistics from the CCSS, the average hospitalization duration in regional hospitals in 2019 due to respiratory diseases was 5.97 days \cite{CCSS2023}. For a visual representation of the geographic distribution of these regions, refer to Figure \ref{fig:mapa}.

\subsubsection{Input data: Climate, pollution, and lagged hospital discharge}

Daily precipitation estimates from 2000 to 2019, sourced from the Climate Hazards Group InfraRed Precipitation with Station data (CHIRPs) \citep{FunkChris2015Tchi}, were used to measure the land surface rainfall. Simultaneously, the Geophysical Research Center (CIGEFI) of the University of Costa Rica supplied maximum and minimum daily temperatures across the country, employing the same estimation procedure as the Climate Hazards Center Infrared Temperature with Stations (CHIRTS) \citep{Funk2019}, with an extension of the time period until 2019. Both data sources offer high spatial resolution at 5km by 5km.

After reviewing the 27 core indices\footnote{See 
\url{https://etccdi.pacificclimate.org/list_27_indices.shtml}} related to climate extremes, as recommended by the  
Climate and Ocean: Variability, Predictability and Change (CLIVAR)\footnote{
\url{https://www.clivar.org/}} on Climate Change Detection \cite{Karl1999},
we adapted them for tropical countries and computed 22 climate change indices for each region related to cumulative rainfall estimates and temperature for each region $i=1,...,7$ and week $t=1,...,T$ (See Table \ref{tab:fitting}).

\begin{table}[htp!]
\caption{\label{tab:fitting} Climate change indices related to rainfall and temperature.}
\begin{tabular}{| m{4.5cm} | m{12cm} |}
\hline
 \begin{center} Variable  \end{center}           &  \begin{center}  Description \end{center}  \\ 
 \hline
1. Tmax\_max            & Maximum (across the region) of maximum   weekly temperature.                                                                                              \\ \hline
2. Tmax\_mean           & Average (across the region) of   maximum weekly temperature.                                                                                              \\ \hline
3. Tmax\_min            & Minimum (across the region) of   maximum weekly temperature.                                                                                              \\ \hline
4. n\_Tmax\_Q3          & Average (across the region) of the   number of days in which the maximum temperature is higher than the percentile 75 of maximum temperature.  \\ \hline
5.  Tmin\_min            & Minimum (across the region) of   minimum weekly temperature.                                                                                              \\ \hline
6.  Tmin\_mean           & Average (across the region) of   minimum weekly temperature.                                                                                               \\ \hline
7.  Tmin\_max            & Maximum (across the region) of   minimum weekly temperature.                                                                                               \\ \hline
8.  n\_Tmin\_Q1          & Average (across the region) of the number of days in which the minimum   temperature is lower than the percentile 25 of minimum temperature.              \\ \hline
9.  amplitude\_max\_max  & Maximum of the maximum daily amplitude of temperature (difference of   daily maximum and minimum temperature).                                            \\ \hline
10.  amplitude\_max\_mean & Average of the maximum daily   amplitude of temperature.                                                                                                  \\ \hline
11.  amplitude\_max\_min  & Minimum of the maximum daily amplitude of temperature.                                                                                                    \\ \hline
12.  amplitude\_min\_max  & Maximum of the minimum daily amplitude of temperature.                                                                                                    \\ \hline
13.  amplitude\_min\_mean & Average of the minimum daily   amplitude of temperature.                                                                                                  \\ \hline
14.  amplitude\_min\_min  & Minimum of the minimum daily   amplitude of temperature).                                                                                                 \\ \hline
15.  n\_amplitude\_Q3     & Average of the number of days in   which the amplitude of temperature is higher than the percentile 75 of daily   amplitude of temperature.               \\ \hline
16.  n\_amplitude\_P90    & Average of the number of days in   which the amplitude of temperature is lower than the percentile 25 of daily   amplitude of temperature.                \\ \hline
17.  precip\_max\_max     & Maximum (across the region) of the   maximum daily precipitation.                                                                                          \\ \hline
18.  precip\_max\_mean    & Average (across the region) of the   maximum daily precipitation.                                                                                          \\ \hline
19.  precip\_max\_min     & Minimum (across the region) of the   maximum daily precipitation.                                                                                         \\ \hline
20.  precip\_mean\_mean   & Average (across the region) of the   average daily precipitation Average (across the region) of the average daily   precipitation.                         \\ \hline
21.  n\_precip\_max\_Q3   & Average (across the region) of the   number of days in which the maximum precipitation is higher than the   percentile 75 of daily maximum precipitation. \\ \hline
22.  n\_precip\_max\_P90  & Average (across the region) of the   number of days in which the maximum precipitation is higher than the   percentile 90 of daily maximum precipitation. \\
\bottomrule
\end{tabular}
\end{table}

In relation to pollution data, daily measurements of AOD from 2000 to 2019 were obtained from the MODIS Atmosphere L3 Daily Product (MOD08\_D3) \cite{MODIS2017} from the National Aeronautics and Space Administration (NASA). We obtained daily data points at 12 locations across the country using a spatial resolution of $1 \times 1$ degree grid. Subsequently, we calculated the weighted average based on the Euclidean distance between the 12 data points and the centroid of each region $i$. Finally, weekly AOD values were derived by averaging the daily information.

All precipitation indices were logarithmically transformed due to their asymmetry. Furthermore, we computed the sample cross-correlation function between all covariates and hospital discharges. We observed that the highest correlation between all covariates and hospital discharges occurs in the same week (\textit{i.e.}, at $0$ lag), except for AOD, where the highest sample cross-correlation is detected with a lag of $10$ weeks, namely AOD\_10, thus we incorporate this lagged covariate instead of the same-week AOD in the model. Finally, hospital discharges from both the previous week and the week before are used in the model.

\subsection{Statistical Methods}

We employed a supervised statistical learning approach to forecast hospital discharges based on the environmental information in each region. The prediction model for hospital discharges at time $t$ ($HD_t$) in a fixed region $i$ is defined as follows:

$$HD_t \sim f(\boldsymbol{X}_t, AOD_{t-10}, HD_{t-1}, HD_{t-2}),$$
where $f(\cdot)$ represents the applied machine learning technique, $\boldsymbol{X}_t$ encompasses all climate covariates at time $t$, and $AOD_{t-10}$ denotes the aerosol optical depth at time $t-10$.

For $f(\cdot)$, we apply four different machine learning methods:
\begin{description}
\item [Random Forest](RF) \cite{BreimanLeo2001RF, Hastie2009} is a bootstrapped ensemble method that combines results of regression trees to improve the prediction. 
\item [XGBoost] (XGBOOST) \cite{chen2016_xgboost} is a gradient-boosting algorithm that employs bagging and trains multiple decision trees in order to produce forecasts.
\item[Facebook's Prophet Forecasting Model] (PROPHET) \cite{Taylor2018} is an additive-based model that fits non-linear trends with multiple seasonality plus holiday factors.
\item [Ensemble method] (ENSEMBLE) is a combination of the predictions of the three aforementioned models with equal weights.
\end{description}

To train the model, we split the data into training (2001-2018) and testing sets (2019).  The training process involves a systematic approach, where the training set is divided into sliding windows. Each sliding window comprises a 2-year analysis set and a 1-year assessment set, accounting for the temporal annual seasonality inherent in the data. After that, we perform model tuning using the grid search method and minimize the root mean squared error (RMSE). Once we obtain the optimized hyperparameters (see Table \ref{table-hyper} in the Supplementary section), we predict hospital discharges and their prediction interval using conformal inference \cite{Lei2018_conformal} in 2019 to compare with the testing set. The use of conformal inference ensures that prediction intervals are free from any assumptions about the distribution of our outcome, and it also allows for achieving interesting coverage properties even for small samples (see \cite{Lei2018_conformal}).

Finally, we compute three metrics to compare the predictive performance of the applied methods for a fixed region. The first one, the Mean Squared Error ($MSE$) is defined as follows:
\begin{align*}
	MSE = \frac{1}{m}\sum_{t=1}^m(HD_t-\widehat {HD}_t)^2,
\end{align*}
where $m$ is the number of weeks in the testing period, $HD_t$ is the observed hospital discharge at week $t$, and $\widehat{HD}_t$ is the estimated hospital discharge according to each model at time $t$. The Mean Absolute Error ($MAE$) is computed as follows:
\begin{align*}
	MAE = \frac{1}{m}\sum_{t=1}^m |HD_t-\widehat {HD}_t)|,
\end{align*}
and finally, the Interval Score (IS) \cite{Gneiting2007a} is defined as 
\begin{align*}
	IS_{\alpha}=\frac{1}{m} \sum_{t=1}^m\left[(U_t-L_t)+\frac{2}{1-\alpha}(L_t-HD_t)\cdot 1_{HD_t<L_t}+\frac{2}{1-\alpha}(HD_t-U_t)\cdot 1_{HD_t>U_t}\right],
\end{align*}
where $U_t$ and $L_t$ are the upper and lower limits of the prediction interval of $(1-\alpha) \%$, respectively.  

$MSE$ and $MAE$ evaluate the point forecasts with the observed hospital discharges. On the other hand, $IS_{\alpha}$ assesses a $(1- \alpha) \%$-prediction interval by comparing its upper and lower limits against the observed values. Finally, to quantify the degree of overfitting in each of the models, we compare the $MSE$ and the $IS$ obtained in the training set versus those obtained in the testing period through the relative $MSE$ and relative $IS$ as follows:

$$
MSE_{rel}=\frac{MSE_{test}}{MSE_{train}}, ~~~~~\text{and} ~~~~ IS_{rel}=\frac{IS_{test}}{IS_{train}}.
$$

We also calculate the contribution of each covariate in predicting the dependent variable. In particular, we are interested in quantifying the global contribution through the aggregation of local contributions based on Shapley values \cite[for more details, see][]{Molnar2020,Strumbelj2014}. This allows us to compare which covariates have a greater impact on the prediction process for each specific region.

All calculations were performed in the statistical software R \cite{R2023} and the following packages \verb|tidymodels| \cite{tidymodels}, \verb|modeltime| \cite{modeltime}, \verb|ranger| \cite{ranger} , \verb|xgboost| \cite{xgboost} \verb|modeltime.ensemble| \cite{modeltime.ensemble}, and \verb|prophet| \cite{prophet}.

\section{Results}
\label{sec:results}
In Table \ref{tablametricas_todos}, we provide a comparison of the four employed methods using the metrics detailed in Section \ref{sec:methods}. It is crucial to highlight that our objective is to assess the predictive performance of the four models during the testing period. This assessment involves comparing the observed outcomes with the projected ones, utilizing the set of covariates observed within the same period.

\begin{table}[htp!]
\caption{Metric comparison by region and method (all covariates). The best method for each region is shown in bold.}
\label{tablametricas_todos}
\centering
{\small
\begin{tabular}{llrrrrrrrr}
  \toprule
Region & Model & $MSE$ & $MAE$ & $IS_{0.05}$ & $MSE$(train) & $MAE$(train) & $IS_{0.05}$(train) & $MSE_{rel}$ & $IS_{rel}$ \\ 
  \midrule
\multirow{3}{*}{Brunca} & RF & 188.55 & 10.18 & 71.79 & 73.10 & 7.00 & 60.09 & 68.78 & 83.69 \\ 
   & XGBOOST & 253.34 & 11.74 & 79.12 & 74.65 & 6.94 & 69.71 & 59.13 & 88.10 \\ 
   & \textbf{PROPHET} & 237.86 & 13.29 & 63.68 & 55.41 & 6.08 & 59.29 & 45.77 & 93.10 \\ 
   & ENSEMBLE & 205.36 & 11.29 & 70.83 & 58.74 & 6.25 & 34.67 & 55.32 & 48.94 \\
  \midrule
  \multirow{3}{4em}{Central Norte} & \textbf{RF} & 222.95 & 12.01 & 69.65 & 97.77 & 7.80 & 56.27 & 64.89 & 80.79 \\ 
   & XGBOOST & 281.86 & 13.70 & 75.23 & 0.00 & 0.03 & 71.80 & 0.21 & 95.44 \\ 
   & PROPHET & 2947.93 & 48.38 & 210.24 & 99.48 & 7.80 & 191.61 & 16.12 & 91.14 \\ 
   & ENSEMBLE & 540.34 & 18.11 & 112.19 & 38.07 & 4.89 & 29.89 & 27.00 & 26.64 \\
  \midrule
  \multirow{3}{4em}{Central Sur} & \textbf{RF} & 407.37 & 16.63 & 85.27 & 166.16 & 10.01 & 76.44 & 60.21 & 89.64 \\ 
   & XGBOOST & 429.42 & 17.28 & 86.63 & 0.00 & 0.03 & 79.24 & 0.17 & 91.47 \\ 
   & PROPHET & 660.35 & 21.00 & 110.24 & 131.06 & 9.23 & 89.29 & 43.97 & 81.00 \\ 
   & ENSEMBLE & 408.17 & 17.28 & 89.01 & 57.41 & 6.14 & 35.92 & 35.52 & 40.36 \\
  \midrule
  \multirow{3}{*}{Chorotega} & RF & 90.16 & 7.34 & 44.12 & 24.29 & 3.87 & 41.31 & 52.71 & 93.64 \\ 
   & \textbf{XGBOOST} & 84.14 & 7.11 & 42.12 & 32.86 & 4.49 & 42.01 & 63.18 & 99.75 \\ 
   & PROPHET & 96.64 & 8.09 & 47.00 & 29.61 & 4.22 & 35.05 & 52.15 & 74.58 \\ 
   & ENSEMBLE & 81.11 & 7.18 & 42.19 & 25.60 & 3.94 & 24.81 & 54.92 & 58.81 \\
  \midrule
  \multirow{3}{4em}{Huetar Atlántica} & \textbf{RF} & 142.02 & 8.77 & 58.77 & 32.12 & 4.36 & 57.38 & 49.73 & 97.63 \\ 
   & XGBOOST & 156.60 & 9.13 & 62.70 & 34.71 & 4.50 & 58.33 & 49.28 & 93.03 \\ 
   & PROPHET & 237.39 & 12.56 & 66.79 & 22.09 & 3.72 & 63.98 & 29.61 & 95.79 \\ 
   & ENSEMBLE & 160.25 & 9.61 & 61.59 & 25.67 & 3.90 & 24.55 & 40.55 & 39.87 \\
  \midrule
  \multirow{3}{4em}{Huetar Norte} & RF & 158.97 & 10.13 & 55.13 & 27.52 & 4.12 & 49.76 & 40.70 & 90.26 \\ 
   & XGBOOST & 180.60 & 10.85 & 60.19 & 48.21 & 5.42 & 54.86 & 49.94 & 91.13 \\ 
   & \textbf{PROPHET} & 163.56 & 10.15 & 58.24 & 34.34 & 4.72 & 55.85 & 46.52 & 95.89 \\ 
   & ENSEMBLE & 161.28 & 10.20 & 56.61 & 33.23 & 4.51 & 27.50 & 44.21 & 48.58 \\
  \midrule
  \multirow{3}{4em}{Pacífico Central} & RF & 53.64 & 5.89 & 34.90 & 35.48 & 4.73 & 31.01 & 80.21 & 88.87 \\ 
   & XGBOOST & 47.99 & 5.46 & 32.65 & 56.81 & 5.83 & 35.97 & 106.65 & 110.16 \\ 
   & \textbf{PROPHET} & 85.38 & 8.18 & 32.11 & 34.52 & 4.67 & 32.05 & 57.06 & 99.82 \\ 
   & ENSEMBLE & 55.21 & 6.20 & 31.03 & 37.39 & 4.82 & 28.68 & 77.67 & 92.42 \\ 
   \bottomrule
\end{tabular}
}
\end{table}

Note that due to differences in the behavior of the series in each region, it is confirmed that there is no method that is optimal for all cases. Likewise, the ensemble method is not optimal according to the calculated metrics. In order to choose the proposed models, we sought to have the smallest $IS$ in the testing period, and we also ensured an $IS_{rel}$ ratio higher than 80\%, with the goal of selecting alternatives with minimal overfitting. The other metrics were used to verify that the selected methods are also competitive compared to alternatives. In general terms, all selected methods successfully predict in the testing period across the regions and, furthermore, allow capturing both high and low-frequency characteristics in the original discharge series.

\begin{figure}[htp!]
    \centering
    \includegraphics[scale=0.85]{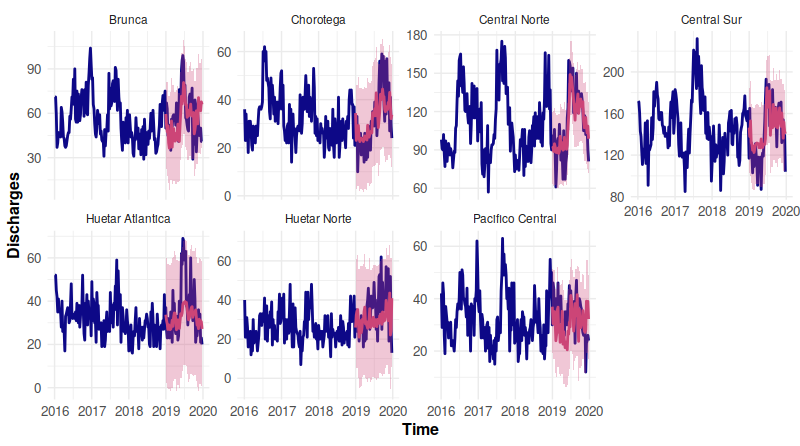}
    \caption{Forecast comparison of hospital discharges, by region.}
    \label{fig:predicciones}
\end{figure}

In Figure \ref{fig:predicciones}, we present these results. Note that in most regions, the 95\% confidence intervals successfully capture the observed series as well. The regions that showed greater difficulty in their forecasts were the Huetar Atlantica and Norte regions, and partially the Pacifico Central region. 

\begin{table}[htp!]
\caption{Relative Importance factors for the best methods (in percentage) by means of the Shapley values. The \textit{date} covariate corresponds to the time index, needed for the PROPHET algorithm.}
\label{tablaimport}
\centering
\begin{tabular}{l|rp{1cm}p{1cm}lp{1.6cm}p{1cm}p{1.5cm}}
  \toprule
\textbf{Covariate} & Brunca & Central Norte & Central Sur & Chorotega & Huetar Atlantica & Huetar Norte & Pacifico Central \\ 
  \midrule
date & 21.1 &  &  &  &  & 18.4 & 23.2 \\ 
  n\_amplitude\_Q3 & 11.0 &  & 2.0 &  & 4.4 & 7.5 & 7.6 \\ 
  n\_amplitude\_P90 & 9.9 &  &  &  & 6.7 &  &  \\ 
  amplitude\_max\_max & 9.9 &  &  &  &  & 6.1 & 7.4 \\ 
  egreso\_2 & 9.3 & 21.6 & 18.9 & 36.1 & 31.0 &  & 8.8 \\ 
  precip\_max\_min & 9.1 &  &  & 3.1 & 3.1 &  & 6.6 \\ 
  precip\_mean\_mean & 9.0 & 3.0 & 3.0 &  &  & 7.7 &  \\ 
  Tmax\_max & 7.9 &  &  &  &  & 13.1 &  \\ 
  n\_precip\_max\_Q3 & 6.6 & 2.3 &  & 3.0 & 3.6 &  &  \\ 
  Tmax\_mean & 6.1 &  & 1.5 &  & 4.0 & 10.9 &  \\ 
  egreso\_1 &  & 46.5 & 50.3 & 16.8 & 24.5 &  &  \\ 
  aerosol\_10 &  & 15.6 & 16.5 & 17.1 & 14.0 &  &  \\ 
  aerosol &  & 4.3 &  & 4.5 &  &  &  \\ 
  precip\_max\_max &  & 2.2 & 1.6 & 3.8 &  &  &  \\ 
  precip\_max\_mean &  & 1.6 & 2.3 &  &  & 6.1 &  \\ 
  Tmin\_min &  & 1.5 &  & 2.8 &  & 7.3 & 7.5 \\ 
  Tmin\_mean &  & 1.4 &  &  &  & 13.8 & 13.0 \\ 
  amplitude\_min\_min &  &  & 2.1 &  &  &  &  \\ 
  amplitude\_min\_mean &  &  & 1.8 &  &  &  & 5.7 \\ 
  amplitude\_max\_min &  &  &  & 9.3 &  & 9.2 &  \\ 
  n\_Tmin\_Q1 &  &  &  & 3.5 &  &  &  \\ 
  amplitude\_max\_mean &  &  &  &  & 5.8 &  & 10.4 \\ 
  amplitude\_min\_max &  &  &  &  & 2.9 &  &  \\ 
  Tmin\_max &  &  &  &  &  &  & 10.0 \\ 
   \bottomrule
\end{tabular}
\end{table}

For these last regions, the PROPHET method constitutes an interesting alternative, since this method also allows incorporating a time index that can explain those non-stationary components not being explained by the covariates. Also, in Table \ref{tablaimport}, we show the Shapley values, all in percentage terms to facilitate interpretation. Note that for those regions where the RF and XGBOOST methods are the best, the covariates with the greatest contribution to the prediction of hospital discharges tend to be the lagged discharge variables (orders 1 and 2) along with the lagged AOD variable. On the other hand, depending on the region, there are environmental variables that contribute to predictions to varying degrees once the contribution of lagged variables has been taken into account. For example, it is noteworthy that for the Chorotega and Huetar Norte regions, the amplitude between maximum and minimum temperature is important, as well as in the latter region, the regional minimum of minimum temperature and the regional maximums of maximum temperature.

In general terms, a considerable number of the covariates we used show significant contribution in the prediction models.

\section{Discussion}
\label{sec:discussion}

To summarize, we used climatic and pollution data to model and predict weekly hospital discharges due to respiratory conditions, a leading contributor to the global healthcare burden in terms of costs, at regional hospital levels in Costa Rica. Four statistical learning approaches: RF, XGBOOST, PROPHET, and the ENSEMBLE method, were applied to each of the seven regions. Furthermore, we compared their predictive capacity in both the training set (2001-2018) and the testing set (2019) using MSE, MAE, and IS, as well as $MSE_{rel}$ and $IS_{rel}$ to assess overfitting. We conclude that the optimal model for each region depends on specific environmental and pollution variables, which do not align for all cases. However, at least for the majority of regions, the lagged variables of hospital discharges and the lagged aerosol variable are of the highest relative importance. In other cases, the non-stationary component is strong enough for a time-indexed model to be preferable. Despite all of the above, the selected model for each region demonstrates its ability to generate reliable forecasts.

As limitations, it is important to note that other significant factors, such as the socio-economic conditions of each region, could serve as potential predictors for hospitalizations due to respiratory conditions. However, many socio-economic factors, including GDP, population density, and healthcare facilities, among others, are not available on a weekly basis or remain constant over time. We argue that, given the established association in the literature between social factors and AOD, we indirectly incorporate this information into the model, recognizing it as a proxy for the anthropogenic effect on climate.

This study enables health authorities to anticipate and visualize the hospital requirements for serving the population with respiratory problems in Costa Rican Health Regions. Additionally, it empowers authorities to administer and allocate limited human and economic resources effectively and efficiently, fostering a balanced national health economy.

\bibliography{reference}  

\section*{Supplementary}

\begin{table}[ht]
\caption{Optimal hyperparameters found using grid search for each method and region. For the RF and XGBOOST methods, the number of trees was held constant at 1000.}
\label{table-hyper}
\centering
{\small
\begin{tabular}{l|l|lp{1cm}p{1cm}lp{1.7cm}p{1.5cm}p{1.5cm}}
  \toprule
\textbf{Method} & \textbf{Hyperparameter} & Brunca & Central Norte & Central Sur & Chorotega & Huetar Atlantica & Huetar Norte & Pacifico Central \\ 
  \midrule
\multirow{2}{*}{RF} & mtry & 25 & 20 & 17 & 15 & 12 & 20 & 22 \\  
   & min\_n & 40 & 22 & 30 & 30 & 40 & 40 & 37 \\ 
   \midrule
  \multirow{6}{*}{XGBOOST} & mtry & 12 & 14 & 14 & 12 & 12 & 10 & 10 \\ 
   & min\_n & 40 & 8 & 8 & 40 & 40 & 35 & 35 \\ 
   & tree\_depth & 13 & 11 & 11 & 13 & 13 & 12 & 12 \\ 
   & learn\_rate & 0.007 & 0.078 & 0.078 & 0.007 & 0.007 & 0.003 & 0.003 \\ 
   & loss\_reduction & 0.001 & 0.018 & 0.018 & 0.001 & 0.001 & 0.000 & 0.000 \\ 
   & sample\_size & 0.898 & 0.828 & 0.828 & 0.898 & 0.898 & 0.589 & 0.589 \\
   \midrule
  \multirow{2}{*}{PROPHET} & prior\_scale\_changepoints & 1.000 & 1.778 & 1.000 & 1.000 & 3.162 & 1.778 & 1.000 \\ 
   & prior\_scale\_seasonality & 3.162 & 1.778 & 10.000 & 1.000 & 1.778 & 5.623 & 1.000 \\ 
   \bottomrule
\end{tabular}
}
\end{table}

\end{document}